\documentclass[aps,prl,twocolumn,groupedaddress]{revtex4}
\usepackage{graphicx}
\input epsf
\usepackage{amsmath,amssymb}
\begin{document}
\title{Blocking transition of SrTiO$_3$ surface dipoles in MoS$_2$/SrTiO$_3$ field effect transistors with counterclockwise hysteresis}
\author{Santu Prasad Jana}
\affiliation{Department of Physics, Indian Institute of Technology Kanpur, Kanpur 208016, India}
\author{S Sreesanker}
\affiliation{Department of Physics, Indian Institute of Technology Kanpur, Kanpur 208016, India}
\author{Suraina Gupta}
\affiliation{Department of Physics, Indian Institute of Technology Kanpur, Kanpur 208016, India}
\author{Anjan K. Gupta}
\affiliation{Department of Physics, Indian Institute of Technology Kanpur, Kanpur 208016, India}
\date{\today}
\begin{abstract}
A counterclockwise hysteresis is observed at room temperature in the transfer characteristics of SrTiO$_3$ (STO) gated MoS$_2$ field effect transistor (FET) and attributed to bistable dipoles on the STO surface. The hysteresis is expectedly found to increase with increasing range, as well as decreasing rate, of the gate-voltage sweep. The hysteresis peaks near 350 K while the transconductance rises with rising temperature above the room temperature. This is attributed to a blocking transition arising from an interplay of thermal energy and an energy-barrier that separates the two dipole states. The dipoles are discussed in terms of the displacement of the puckered oxygen ions at the STO surface. Finally, the blocking enables a control on the threshold gate-voltage of the FET over a wide range at low temperature which demonstrates it as a heat assisted memory device.
\end{abstract}
\maketitle
 \section{Introduction}
Atomically thin semiconducting transition metal dichalchogenides (TMDs) \cite{TMDs} are advantageous when integrated into field effect transistors (FETs). Molybdenum disulfide (MoS$_{2}$) is a convenient TMD for probing and optimizing such two dimensional (2D) material devices due to its natural abundance and environmental stability. It also offers mechanical flexibility, high transparency, layer-thickness-dependent bandgap \cite{direct gap,direct gap1}, higher mobility values than organic semiconductor FETs and excellent electrostatic gate control. Many electronic \cite{direct gap1,logic,logic1,high frequency} and optoelectronic \cite{photodetectors,direct gap1,light,photo1} applications have been successfully demonstrated using few layer MoS$_2$.

\par Interface-traps between a 2D semiconductor and dielectric gate or the substrate, in a FET, pose a challenge for 2D material electronic devices as they lead to a clockwise (or positive) hysteresis in the transfer characteristics and reduced mobility which limit the devices' potential \cite{MoS2 2012,MoS2 20161,MoS2 20162,Mos2 2023,Ta2O5}. The hysteresis has also sparked interest in the potential applications of MoS$_{2}$ in thermally assisted non-volatile memory devices \cite{Ta2O5}. In this context, integrating MoS$_{2}$ with a ferroelectric (FE) substrate, such as Lead-zirconate-titanate (PZT), is promising as a FE offers higher polarization induced charge density than a dielectric substrate. A PZT substrate in MoS$_{2}$ FETs often shows the expected counterclockwise (or negative) hysteresis \cite{counterclockwise,counterclockwise1,both}. However, an unwanted ``anti-hysteresis", similar to that due to traps, has also been reported \cite{clockwise,clockwise1,clockwise2} due to accumulation of charges at the polar ferroelectric surface that leads to polarization screening.

\par High dielectric constant gate materials, such as SrTiO$_3$ (STO) \cite{STO5}, are also of interest as they help suppress long-range charged impurity scattering, thereby enhancing the carrier mobility. A number of studies have investigated transport in graphene on STO thin films \cite{GSTO film,GSTO film1} and STO bulk substrates \cite{GSTO bulk,GSTO both,GSTO bulk1,GSTO bulk4} over a wide temperature range to examine the effects of its large and tunable dielectric constant. Although STO is widely recognized as a paraelectric \cite{1973,1976,STO6,STO7,STO-SHG} material, a rare ferroelectric-like counter-clockwise hysteresis \cite{GSTO both} has been reported in graphene-STO FETs and attributed to puckered oxygen ions on the STO surface together with the more common clockwise hysteresis \cite{GSTO film,GSTO film1,GSTO bulk1, GSTO bulk4}.

\par In this paper, we report on ferroelectric-like counterclockwise hysteresis in MoS$_{2}$ FET with STO single-crystal substrate as back-gate. The hysteresis increases with increasing sweep-range and with decreasing sweep-rate of the gate voltage. With temperature increasing above room temperature, the transconductance monotonically rises but the hysteresis peaks near 350 K. This is attributed to the bistability of the puckered oxygen ions on the STO surface enabling two dipole states with gate electric field providing a bias between the two states. The hysteresis peak is discussed in terms of the blocking transition of the surface dipoles. The blocking is used to program the dipoles' state through gate-cooling to control the FET's threshold voltage at low temperatures.

\section{Experimental Details}
Acetone/IPA cleaned 0.5 mm thick single side polished STO substrate (from Crystal Gmbh) with (001) orientation is used as gate dielectric. A 50 nm thick gold film deposited by thermal evaporation on the rough surface of STO substrate, serves as the backgate electrode. Multi-layer MoS$_{2}$ flake is mechanically exfoliated from natural bulk crystal (from SPI) and transferred on the STO by a dry transfer method \cite{XYZ} using PDMS film (Gel film from Gel Pak) and an XYZ micromanipulator attached with an optical microscope. To increase the adhesion between the MoS$_{2}$ flake and the STO substrate, hexamethyldisilane (HMDS) was spin-coated at 2000 rpm for 45 s prior to the MoS$_{2}$ transfer on the polished side of the STO substrate. This also helps in eliminating the effect of interface traps \cite{HMDS}. Nevertheless, the blocking transition is found to be unaffected by the presence of the HMDS layer or by the roughness of the STO surface, see Appendix.
\begin{figure}[h]
	\centering
 	\includegraphics[width=3.4in]{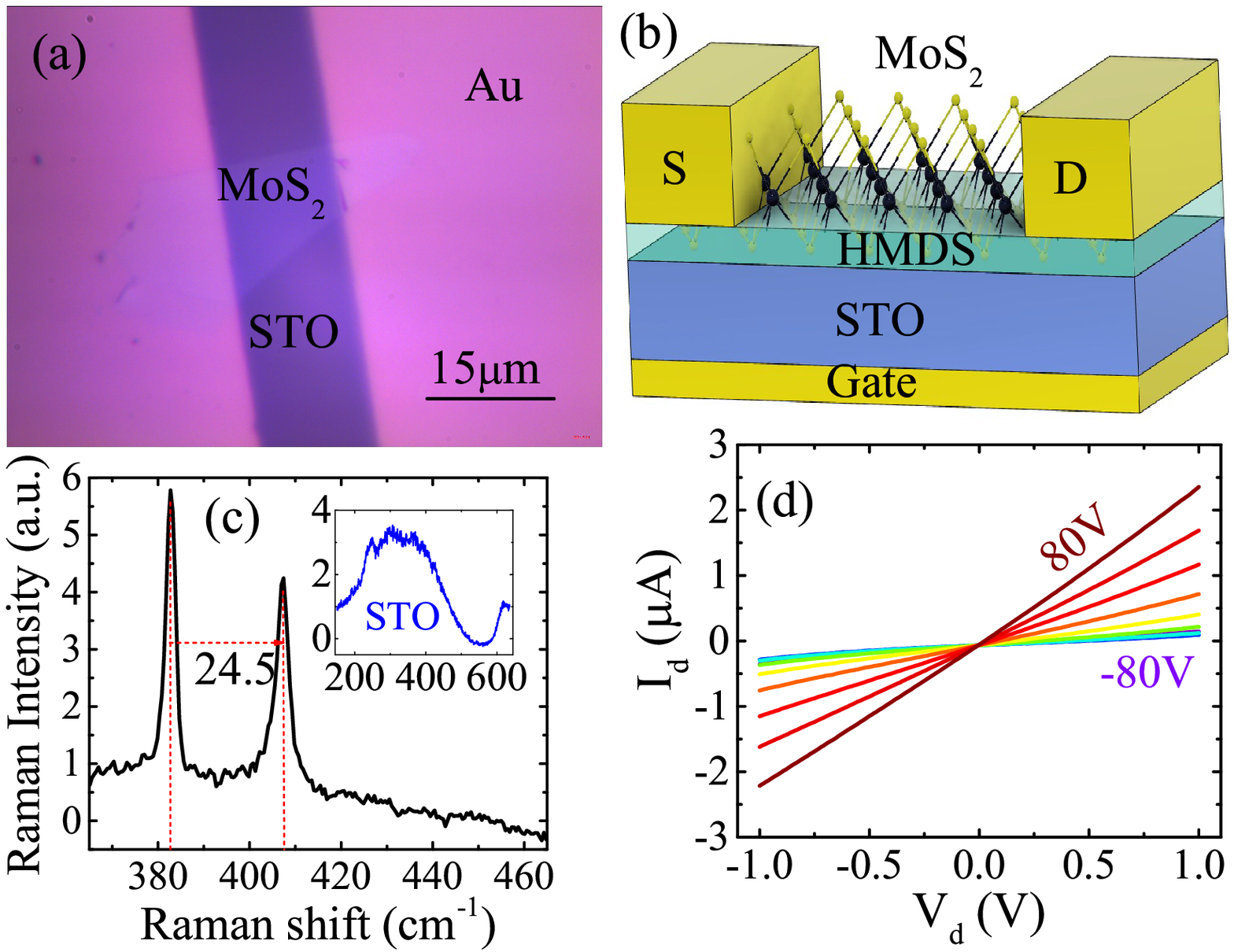}
	\caption{(a) Optical image of multi-layer MoS$_2$ on HMDS coated STO substrate with gold contacts. (b) shows a schematic illustration of the FET device with source drain and gate electrodes of gold. (c) Raman spectrum of exfoliated multi-layer MoS$_2$ on STO. The inset shows the Raman spectrum of the bare STO substrate. (d) $I_{\rm d}$ versus $V_{\rm d}$ curves of the device at room temperature at different $V_{\rm g}$ values between -80 and 80 V separated by 20 V.}
	\label{fig:GSTO1}
\end{figure}

Source and drain contacts of gold are made using mechanical masking with a 15 $\mu$m diameter tungsten wire by aligning the MoS$_{2}$ flake underneath the wire with the help of an optical microscope, followed by 50-nm-thick gold film deposition by thermal evaporation. Fig. \ref{fig:GSTO1}(a) shows the optical image of fabricated Multi-layer MoS$_2$/STO FET with gold contacts. Use of mechanical masking helps avoid the organic lithography resists which can leave residue on MoS$_{2}$. Fig. \ref{fig:GSTO1}(b) shows the schematic of the device with source-drain contacts on MoS$_{2}$ as well as gate electrode of Au. Fig. \ref{fig:GSTO1}(c) shows the Raman spectra of the MoS$_2$ with a smooth background coming from the STO substrate. The two characteristic Raman peaks of MoS$_2$, $E^1_{2g}$ and $A_{1g}$, are separated by 24.5cm $^{-1}$ which corresponds to multilayer MoS$_2$ as reported in the literature \cite{Raman,anomalous lattice vibration}. The inset of Fig. \ref{fig:GSTO1}(c) shows the Raman spectra of the bare STO substrate with the second-order vibrational modes of STO \cite{Petzelt}.

\par The two-probe electrical transport measurements used a drain-source voltage bias $V_{\rm d}$ controlled by a data acquisition card. The drain current $I_{\rm d}$ is measured through the voltage across a small bias resistor in series with the channel and by using a differential voltage amplifier. A 10 k$\Omega$ series resistor was connected to the gate voltage $V_{\rm g}$ supply to protect it and other electronics in case of gate failure. The gate supply was controlled by the data acquisition card using a LabView program. The linear two-probe current-voltage characteristics, illustrated in Fig. \ref{fig:GSTO1}(d) at different $V_{\rm g}$ values, confirmed the Ohmic contacts. All electrical measurements down to 80 K were conducted in a homemade vacuum cryostat with a heater for temperature control. The cryostat was pumped by a turbomolecular pump to better than 10$^{\rm-4}$ mbar vacuum which improves further when the cryostat is immersed in liquid nitrogen for cooling.

\section{Blocking transition in bistable system}
Here, basic physics of blocking transition, relevant for the studied MoS$_2$/STO FET device, is reviewed briefly. A bistable system consists of two metastable minimum energy states that are separated by an energy barrier of height $\Delta_0$, see the free energy plot in Fig. \ref{fig:GSTO2}(a). A biasing energy $E_{\rm B}$, enabled by the gate electric field in the FET, makes the heights of the barrier, as seen from the two minima, unequal, \emph{i.e.} $\Delta_0\pm E_{\rm B}$. This leads to more thermally activated transitions from the shallow to the deep minimum than the other way and larger equilibrium population in the deep minimum. Note that we are assuming an ensemble of identical such two state systems. A superparamagnetic nanoparticle is an example of such a bistable system, see suppl-info of ref-\cite{Mos2 2023}, which exhibits blocking transition. Here, the barrier arises from the magnetic anisotropy energy and the biasing energy is dictated by the applied magnetic field.

At a given temperature $T$, the transition rates $\tau_{1,2}^{-1}$ between the two minima are given by $\tau_0^{-1}\exp[-(\Delta_0\pm E_{\rm B})/k_{\rm B}T]$ with $\tau_0^{-1}$ as the attempt rate. These rates are monotonically rising with increasing temperature and imply a rise in system's response with temperature for a fixed bias. The interplay of the transition rate and the rate $\tau_{\rm m}^{-1}$ of the experimentally controlled bias-energy change, and the history, dictate the evolution of the population distribution in the two minima. The transition rates $\tau_{1,2}^{-1}$ are steeply rising functions of temperature and these will match with $\tau_{\rm m}^{-1}$ at some temperature $T_{\rm B}$ given by, $T_{\rm B}\approx\Delta_0/[k_{\rm B}\ln(\tau_{\rm m}/\tau_0)]$. Note that a very small bias, \emph{i.e.} $E_{\rm b}\ll\Delta_0$, is assumed so that it does not make any of the minima disappear and the transition rates are predominantly dictated by $\Delta_0$.

Below $T_{\rm B}$ the system's response to the change in bias will be small while above $T_{\rm B}$ the system will show a large response. The crossover that occurs at $T_{\rm B}$ is called the Blocking transition. Further, when the bias is swept at a constant rate between two opposite extreme biases $\pm E_{\rm b0}$ over a time $\tau_{\rm m}$, the population-difference between the two minima at zero bias during the two sweep directions can be written as
\begin{align}
\Delta p \propto \left[1-\exp\left\{-\frac{\tau_{\rm m}}{2\tau_0}\exp\left(-\frac{\Delta_0 -aE_{\rm b0}}{k_{\rm B}T}\right)\right\}\right]&\nonumber\\
\times \exp\left\{-\frac{\tau_{\rm m}}{2\tau_0}\exp\left(-\frac{\Delta_0 +bE_{\rm b0}}{k_{\rm B}T}\right)\right\}&.
\label{eq:sup-par-analytical}
\end{align}
Here, $a$ and $b$ are empirical constants less than one and depend on the magnitude of the bias. This expression has been elaborated in the suppl-info of ref-\onlinecite{Mos2 2023} for a superparamagnetic system.

\section{Results and analysis}
\begin{figure}[h]
\centering
\includegraphics[width=3.4in]{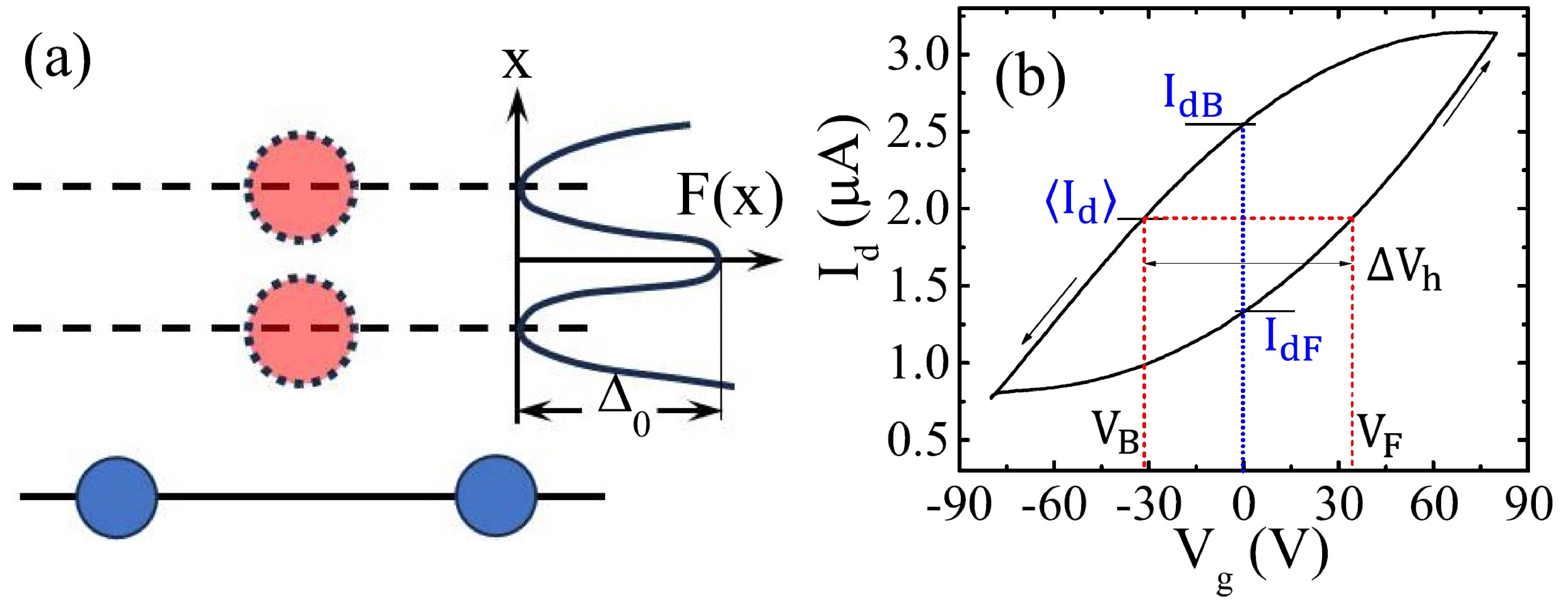}
\caption{The schematic (a) depicts the two minimum free-energy [$F(x)$] states of the puckered oxygen ions (red spheres) at the STO surface with large and small displacements ($x$) from the positively charged Sr/Ti ions (blue spheres). A negative $V_{\rm g}$ lowers the energy of the state with large displacement having larger inward dipole-moment than the other state. (b) shows the measured $V_{\rm g}$-dependence of $I_{\rm d}$ for the MoS$_2$/STO FET at V$_{\rm d} = 1$ V over a V$_{\rm g}$ cycle between -80 and 80 V. $I_{\rm dF}$ and $I_{\rm dB}$ are the drain currents at $V_{\rm g}=0$ during forward and backward $V_{\rm g}$-sweeps, respectively. $\Delta V_{\rm h}=V_{\rm B}-V_{\rm F}$ with $V_{\rm F}$ and $V_{\rm B}$ as the $V_{\rm g}$ values at $I_{\rm d}=\langle I_{\rm d}\rangle =(I_{\rm dF}+I_{\rm dB})/2$ for the respective $V_{\rm g}$-sweep directions.}
\label{fig:GSTO2}
\end{figure}

Figure \ref{fig:GSTO2}(b) shows the room temperature $I_{\rm d}-V_{\rm g}$ characteristics of multi-layer MoS$_2$-STO FET at $V_{\rm d}=1$ V. The forward $V_{\rm g}$ sweep curve is seen to be shifted towards right while the reverse-sweep curve is shifted towards left. This amounts to a counterclockwise hysteresis as opposed to the clockwise hysteresis in MoS$_2$/SiO$_2$ FETs \cite{Mos2 2023} due to the interface traps. In these MoS$_2$-STO FETs, the threshold voltage $V_{\rm th}$ is not seen in all the $I_{\rm d}-V_{\rm g}$ curves over used $V_{\rm g}$ range. Thus, we quantify the hysteresis by the apparent relative $V_{\rm g}$ shift, \emph{i.e.} $\Delta V_{\rm h}$, between the two sweep directions. More specifically, as defined in Fig. \ref{fig:GSTO2}(b), $\Delta V_{\rm h}$ is the difference between the reverse and forward $V_{\rm g}$ values at the average of the forward and reverse $V_{\rm g}$-sweep currents at $V_{\rm g}=0$ V.

Such counterclockwise hysteresis is commonly attributed to the ferroelectric (FE) properties of the gate material. In an n-FET with FE gate, a positive $V_{\rm g}$ amounts to an electric field pointing towards the channel leading to FE dipoles' alignment parallel to the field. Now when $V_{\rm g}$ is made zero, a FE remnant state would persist with a polarization pointing towards the channel. In such zero electric field state, the channel will acquire a negative charge density and the gate electrode will have an equal magnitude positive charge density. This to nullify the electric field in the FE due to the remnant polarization. This amounts to electron accumulation in the n-FET channel leading to enhanced conductance and a reduced threshold $V_{\rm g}$ as compared to the zero polarization state of the FE. Similarly, when one arrives at $V_{\rm g}=0$ state from a negative $V_{\rm g}$, there will be depletion of channel electrons resulting into loss in channel conductance and an increase of the threshold $V_{\rm g}$. This is consistent with the counterclockwise hysteresis observed in Fig. \ref{fig:GSTO2}(b).

STO is actually a paraelectric down to $T=0$ K \cite{1973,1976,STO6,STO7} with an inclination towards a ferroelectric transition as the temperature approaches 0 K but in the bulk there is no ferroelectricity in STO at non-zero temperatures. However, the surface atoms experience different boundary conditions and bonding environments compared to the bulk. Modifications in atomic force constants and changed phonon characteristics at the surface may induce a FE reconstruction near the surface at finite temperatures. Observation of second harmonic generation \cite{STO-SHG} shows an inversion symmetry breaking, implying dipoles on the surface of reduced STO. Some loss of oxygen at the exposed surface of an untreated STO cannot be ruled out although extremely sensitive probes may be needed to actually establish it. Further, the theory-experiment correlations suggest that the surface comprises of domains with two distinct layer-terminations. One termination features an oxygen and a strontium atoms per unit cell, \emph{i.e.} O-Sr, referred to as “Sr termination". The other features two oxygen and one titanium atoms per unit cell, \emph{i.e.} O-Ti-O, referred to as “Ti termination" \cite{STO1}. In both terminations, the oxygen atoms get pulled out, or diplaced, from the surface under simultaneous multilayer relaxations \cite{STO1,STO2,STO3,STO4}. This phenomenon, known as “puckering", leads to the formation of a static inward dipole moment on the STO surface. Consequently, a thin surface layer can exhibit permanent dipole moments or a ferroelectric-like state. Such displacement will also couple to the electric field.

We believe that this FE-like counterclockwise hysteresis in the MoS$_2$-STO FETs arises from the response, or the dynamics, of the STO-surface oxygen-ion displacement to the gate electric field. Presence of electric field near STO surface will promote displacement of puckered oxygen ions in a direction opposite to the field which increases the dipole moment parallel to the field. This is similar to the FE polarization. Further, from the observed blocking transition, discussed later, such surface dipoles seem to exhibit a bistable behavior. This can arise from two local minima in the energy with respect to the puckered oxygen-ion displacement with an energy barrier between the two, see Fig. \ref{fig:GSTO2}(a). From the sharpness of the blocking transition, it appears that there is a very narrow distribution of barrier energies at play. The applied gate electric-field provides a bias and lowers the energy of one of the two minima. A similar polarization may also occur at the other STO surface, close to the gate electrode, with the bulk of the STO still behaving as a dielectric. The electric field at the STO surface for a fixed $V_{\rm g}$ will depend on the dielectric constant of the bulk STO, which has a significant temperature dependence.

\par
\begin{figure}[h]
	\centering
 	\includegraphics[width=3.4in]{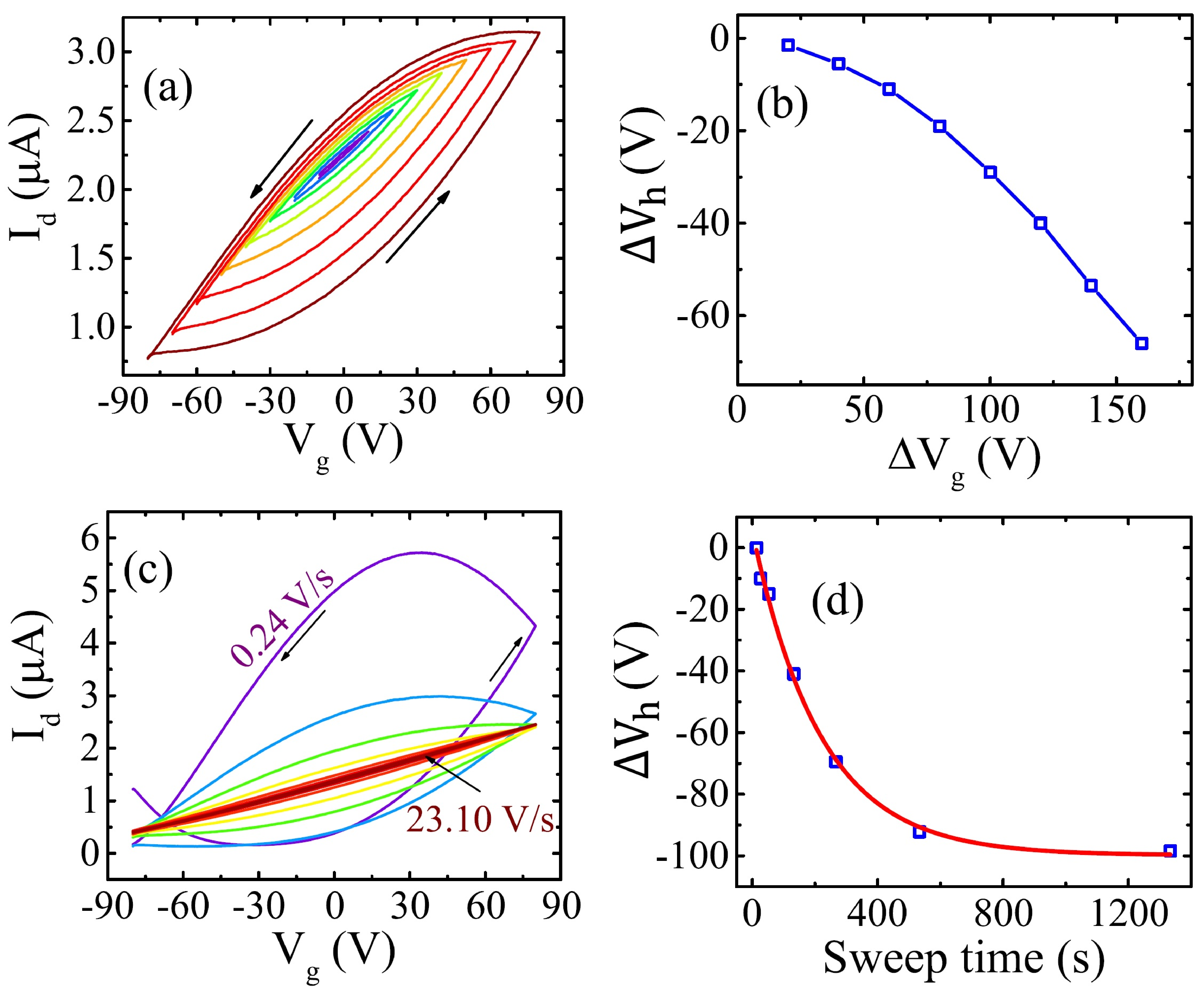}
	\caption{(a) Room temperature $I_{\rm d}-V_{\rm g}$ characteristics at $V_{\rm d}$ = 1 V for different sweep ranges of V$_g$ from $\pm$10 to $\pm$80 V at a constant sweep rate of 1.15 V/s. (b) shows the variation of hysteresis width $\Delta V_{\rm h}$ with $\Delta V_g$. (c)Room temperature $I_{\rm d}-V_{\rm g}$ curves for different sweep rates, between 0.24 and 23.1 V/s, of $V_{\rm g}$ at a fixed $V_{\rm d}$ = 1 V and $\Delta V_{\rm g}=160$ V. (d) shows $\Delta V_{\rm h}$ as a function of total sweep time with the red line showing a fit to a single exponential relaxation with 218 s characteristic time.}
	\label{fig:GSTO3}
\end{figure}

 Figure \ref{fig:GSTO3}(a) shows the measured $I_{\rm d}-V_{\rm g}$ curves for different $V_{\rm g}$ sweep ranges varying from $\pm10$ ($\Delta V_{\rm g} = 20$ V) to $\pm80$ V ($\Delta V_{\rm g} = 160$ V) but at a fixed $V_{\rm g}$ sweep rate of 1.15 V/s. The hysteresis, quantified by $\Delta V_{\rm h}$ increases monotonically with increasing sweep range $\Delta V_{\rm g}$, see Fig. \ref{fig:GSTO3}(b). This happens as more oxygen-ions change their displacement in response to larger $V_{\rm g}$ sweep. As a result the $I_{\rm g}$ response also rises. A non-zero $V_{\rm g}$ acts as a bias leading to a tilt of the bistable potential, see Fig. \ref{fig:GSTO2}(a), and with the tilt proportional to $V_{\rm g}$.
Figure \ref{fig:GSTO3}(c) shows the $I_{\rm d}-V_{\rm g}$ hysteresis loops acquired at different $V_{\rm g}$ sweep rates from 0.24 to 23.1 V/s at room temperature for fixed $V_{\rm g}$-range. Both the hysteresis and the change in $I_{\rm d}$ are seen to reduce with increasing rate. Fig. \ref{fig:GSTO3}(d) shows the dependence of $\Delta V_{\rm h}$ on the overall sweep-time which is inversely related to sweep rate. $\Delta V_{\rm h}$ is found to fit well to an exponential with a characteristic time $\tau_{\rm c}=218$ s. The sweep-range and sweep-rate dependence can be understood from earlier discussion on Blocking transition as the switching rates between the two states will increase with the increase in bias, provided by $V_{\rm g}$, while a smaller sweep rate will enable more transitions for a fixed switching rate of the dipoles.

\begin{figure*}
	\centering
 	\includegraphics[width=6.5in]{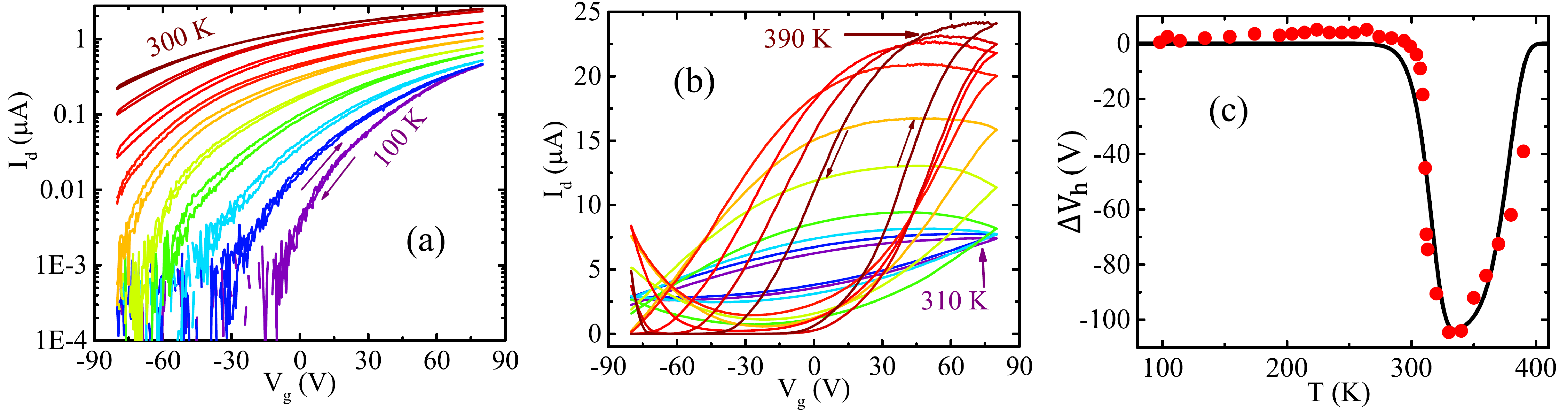}
	\caption{(a) $I_{\rm d}-V_{\rm g}$ curves at $V_{\rm d}$ = 1V and 1.15 V/s sweep rate for different temperatures between 100 and 300 K and (b) between 310 and 390 K. (c) Variation of $\Delta V_{\rm h}$ with temperatures as found from the curves of (a) and (b) with the continuous line as the curve calculated using Eq. \ref{eq:sup-par-analytical} (see text for details).}
	\label{fig:GSTO4}
\end{figure*}
\par Figure \ref{fig:GSTO4}(a,b) display $I_{\rm d}-V_{\rm g}$ curves at $V_{\rm d}=1$ V for temperatures ranging from 80 K to 390 K, over a $V_{\rm g}$ sweep range of $\Delta V_{\rm g}=160$ V at 1.15 V/s sweep rate. The device was first kept at room temperature with $V_{\rm g}=0$ V for over 10 hours to equilibrate and then cooled to 80 K at this gate and then data were acquired after stabilizing different temperatures during the heating and up to 390 K. No difference between cooling and heating runs was found as long as $V_{\rm g}$ was kept at zero during temperature change. The counterclockwise hysteresis observed at room temperature decreases and vanishes with cooling, see Fig. \ref{fig:GSTO4}(a). In fact, the very small hysteresis at low temperatures has an opposite sense, similar to that due to traps, as compared to the one at high temperatures. This could be due to traps either at the MoS$_2$/STO interface or in the bulk of MoS$_2$. The hysteresis, on the other hand, increases with heating above room temperature and peaks near 350 K before decreasing. This can be seen more clearly from the temperature variation of $\Delta V_{\rm h}$ in Fig. \ref{fig:GSTO4}(c).

It can also be seen that the overall $I_{\rm d}$-response to $V_{\rm g}$, representing the transconductance, rises monotonically with increasing temperature above the room temperature, which is in contrast with the non-monotonic behavior of the hysteresis. Since the dielectric constant of bulk STO decreases monotonically with rising temperature one would rather expect a minor reduction in transconductance with rising temperature. From this we infer that the channel's carrier density due to the surface dipoles dominates, at least above the room temperature, over that due to the bulk dielectric polarization of STO. With cooling below room temperature, the response-time of the surface dipoles, to the $V_{\rm g}$-change, grows and becomes much larger than the $V_{\rm g}$ sweep time. This makes the hysteresis vanish and the contribution of surface dipoles to the transconductance will also vanish. With increasing temperature the response time decreases which makes the transconductance increase as more surface dipoles respond to the $V_{\rm g}$ change. The hysteresis peaks at $T_{\rm B}$ where the response-time and the $V_{\rm g}$ sweep-time match. Well above $T_{\rm B}$ nearly all dipoles respond fast and to both, the forward and reverse, sweeps of $V_{\rm g}$. This reduces the hysteresis but increases the transconductance further.

The hysteresis and the blocking transition can be modeled using a bistable system, discussed earlier, for surface dipoles with two energy-minima separated by a barrier $\Delta_0$ with a bias energy $E_{\rm B}$ proportional to $V_{\rm g}$. The continuous line in Fig. \ref{fig:GSTO4}(c) shows a temperature dependent $\Delta V_{\rm h}$ calculated using Eq. \ref{eq:sup-par-analytical} with a single barrier energy $\Delta_0$ and a energy bias $E_{\rm B}$ proportional to $V_{\rm g}$. The parameters used for this calculated curve are $\tau_{\rm m}/\tau_0=6\times10^{14}$, $\Delta_0=1$ eV, $aE_{\rm B0}=bE_{\rm B0}=0.09$ eV at $V_{\rm g}=80$ V. The simple model of blocking transition using a bistable system captures the $\Delta V_{\rm h}$ variation remarkably well.

It is found that the hysteresis together with the blocking transition and the transition temperature is nearly the same in several different studied devices, see Appendix, with varying number of MoS$_2$ layers and MoS$_2$/STO interface. Although the behavior of transconductance below the room temperature somewhat varies between the samples. This could arise from the variation in defect density in MoS$_2$ that can lead to variation in the amount of n-doping together with a rising dielectric constant of STO with cooling. A higher n-doping can lead to larger transconductance assisted by a rising dielectric constant of STO with cooling. This is the case with one of the samples as discussed the Appendix. Anyhow, the surface dipoles' response to $V_{\rm g}$ is expected to diminish with cooling and it is independent of the contribution from other sources such as dopants and the dielectric polarization in the gate.

Although the underlying blocking physics here is similar to that of the interface traps \cite{Mos2 2023} but there are some crucial differences. The activation of traps in response to the gate electric field leads to screening of the electric field seen by the channel while the surface dipoles enhance the field. Thus, in the case of interface traps the sense of hysteresis is opposite, \emph{i.e.} clockwise, and the increasing traps' participation with rising temperature reduces the transconductance or $I_{\rm d}$ change with $V_{\rm g}$. Further, the hysteresis keeps rising with temperature in case of traps with a steep rise at a particular temperature but in case of surface dipoles the hysteresis exhibits a clear peak. This is due to a wide distribution in barrier energy in case of traps but the surface dipoles appear to have a very narrow distribution in barrier energy. This implies a much less disordered nature of surface dipoles in STO than the traps.

\par
\begin{figure}[h]
	\centering
 	\includegraphics[width=3.4in]{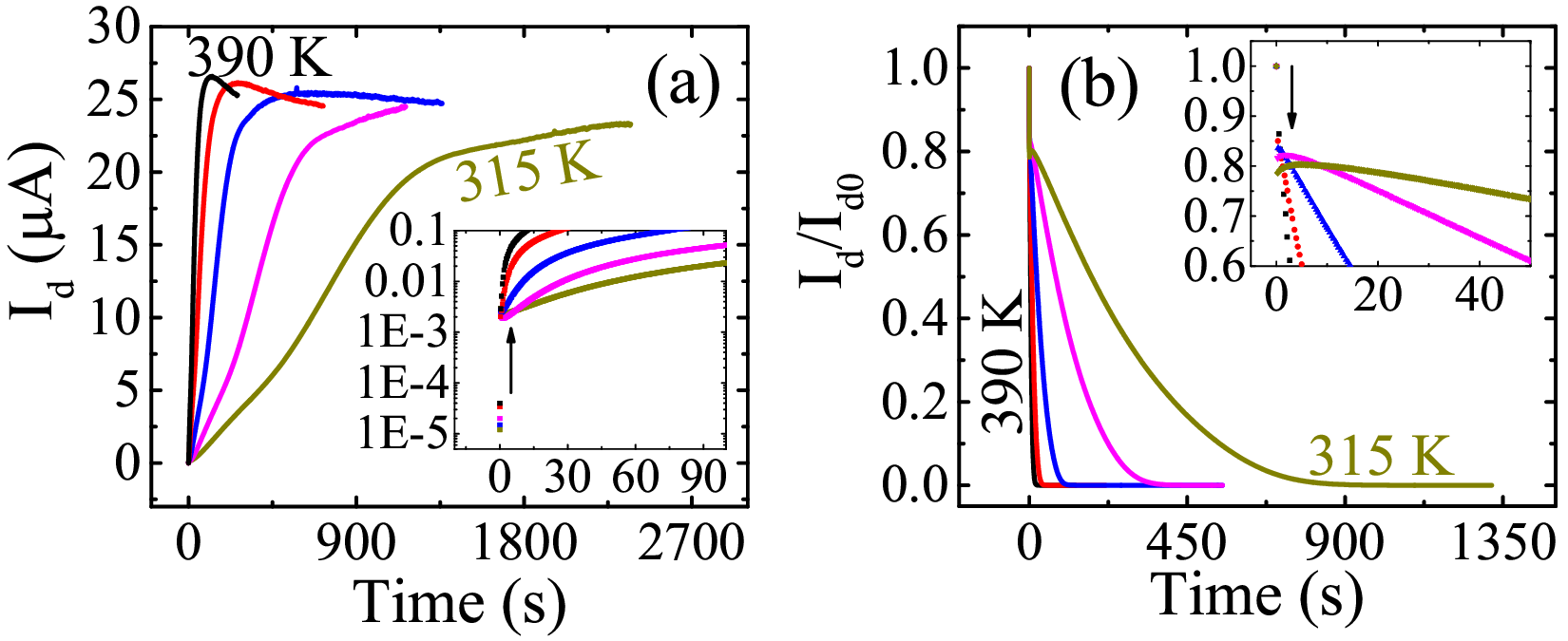}
	\caption{(a) Time-dependence of $I_{\rm d}$ at $T=315, 330, 350, 370$ and 390 K temperatures when $V_{\rm g}$ is changed from -80 to 80 V at $t=0$ s. The inset shows the zoomed-in portion of the graph with the arrow indicating the sudden jump in $I_{\rm d}$ at $t=0$. (b) Time-dependence of $I_{\rm d}$ at $T=315, 330, 350, 370$ and 390 K when $V_{\rm g}$ is stepped from 80 V to -80V at at $t=0$ s. The inset shows the zoomed-in portion with the arrow indicating the sudden jump at $t=0$. In (b), the vertical axis has been normalized with $I_{\rm d0}$ of respective temperatures. $I_{\rm d0}$ is the current at $V_{\rm g}=80$ V and just before the voltage step.}
	\label{fig:GSTO5}
\end{figure}

 Figure \ref{fig:GSTO5}(a) shows the $I_{\rm d}$ relaxations at different temperatures when $V_{\rm g}$ is abruptly switched from -80 to 80 V at $t=0$ s. $V_{\rm g}$ was held at -80 V for more than 3 hours to achieve near equilibrium conditions with the MoS$_2$ channel in the off state. At $t=0$, when the $V_{\rm g}$ is abruptly changed, there is a tiny jump in $I_{\rm d}$ followed by a relatively slow rise. The abrupt jump is attributed to the bulk STO dielectric induced channel carriers while the slow rise is due to the surface dipoles. With increasing temperatures the time associated with the slow rise decreases monotonically while below room temperature this time is extremely large.
Fig. \ref{fig:GSTO5}(b) shows the $I_{\rm d}$ relaxation, scaled with the respective saturated values at 80 V, as a function of time when the $V_{\rm g}$ is changed from 80 to -80 V at $t=0$ s. There is a small abrupt downward jump in $I_{\rm d}$ due to STO dielectric response, see Fig. \ref{fig:GSTO5}(b) inset, followed by a slow decrease due to surface dipoles' response. In both cases, such activated response time behavior is consistent with the two state model discussed earlier.

\par
\begin{figure}[h]
	\centering
 	\includegraphics[width=3.4in]{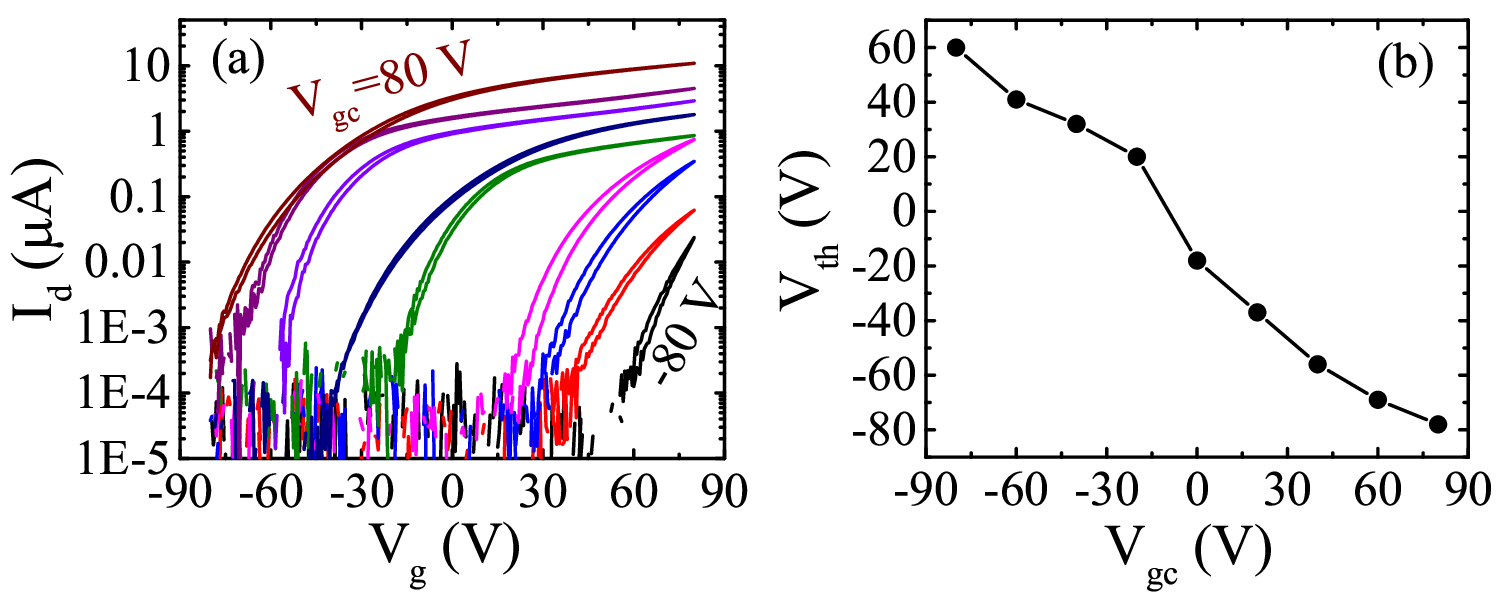}
	\caption{(a) shows the $I_{\rm d}-V_{\rm g}$ curves at $V_{\rm d}=1$ V and at 80 K after cooling the FET from 350 to 80 K under different applied gate voltages $V_{\rm gc}$ from -80 V to 80 V in steps of 20 V. (b) shows the variation of threshold-voltage $V_{\rm th}$ with $V_{\rm gc}$.}
	\label{fig:GSTO6}
\end{figure}

 Figure \ref{fig:GSTO6}(a) shows the effect of cooling on the $I_{\rm d}-V_{\rm g}$ curves measured at 80 K when the device is cooled from 350 to 80 K under different gate-cooling voltages $V_{\rm gc}$ from -80 V to 80 V. For each $V_{\rm gc}$, the device was kept at 350 K in vacuum at the desired $V_{\rm gc}$ for one hour and then cooled to 80 K in its presence. As expected, there is negligible hysteresis at 80 K, but a significant shift in the threshold voltage $V_{\rm th}$ ranging between -78 and +60 V, depending on the $V_{\rm gc}$ value. Fig. \ref{fig:GSTO6}(b) shows the monotonic decrease of $V_{\rm th}$ with $V_{\rm gc}$.
This is easily expected from the blocking transition that happens near 350 K above which the surface dipoles are unblocked and respond fast to the $V_{\rm g}$ change. When gate-cooled, the dipoles get blocked in the changed configuration the dipoles freeze or get blocked again. This enables a reversible control on $V_{\rm th}$. For negative $V_{\rm gc}$, the surface dipoles exhibit a dipole moment pointing away from the channel as compared to zero $V_{\rm gc}$. This depletes electrons from the MoS$_2$ channel and leads to an increase in $V_{\rm th}$. In contrast, the positive $V_{\rm gc}$ leads to a decrease in $V_{\rm th}$. Thus, the STO surface dipoles act as a virtual gate, enabling heat-assisted non-volatile memory.

\section{Discussions and Conclusions}
In a superparamagnet, the magnetic field needed to make the potential monostable is experimentally accessible. As a result, for such large fields the hysteresis is seen at low temperatures but it vanishes at high temperatures and above the blocking transition in contrast with the surface dipoles as well as the interface traps. The field required to make the potential monostable for traps can easily be higher than the breakdown field of the gate material. However, in case of STO surface dipoles, it will be interesting to see if sufficiently large gate electric field is achievable to make the potential monostable. This can have useful consequences for using these surface dipoles in memory applications. It's also noteworthy that MoS$_2$ acts as a sensitive detector of such small ionic displacements.

The counterclockwise hysteresis due to surface dipoles could not be eliminated by a HMDS passivation layer, see Appendix, as opposed to the interface traps \cite{Mos2 2023} where the passivation layer blocks the transfer of charge between the channel and the traps. Thus the presence of HMDS or the roughness, see Appendix, of the STO surface do not eliminate the hysteresis and the blocking transition. Sachs et. al. \cite{GSTO both} found similar counterclockwise hysteresis in graphene-STO FETs at room temperature which vanished with cooling but without any hysteresis in the Au/STO/Au parallel-plate capacitors ruling out ferroelectricity in the bulk STO. A deposited metal layer on STO could alter the surface dipole layer to eliminate their field response while the 2D layer of graphene or MoS$_2$ does not bond enough to alter the surface dipoles.

\par In conclusion, a blocking transition near 350 K resulting from the bistable surface dipoles in STO is found and studied systematically in MoS$_2$/STO FET devices. This transition occurs when the activated transition rate, arising from the interplay of thermal energy and and a barrier-energy between between two states, matches with the gate-voltage sweep-rate. A peak in counterclockwise hysteresis and rising transconductance above room temperature is thus observed. The blocking is used to demonstrate an interesting memory effect of the surface dipoles enabling a control on the FET's threshold voltage at low temperatures which has application potential in non-volatile memory. This also illustrates MoS$_2$ as a sensitive detector of surface polarization.

\section{ACKNOWLEDGMENTS}
The authors acknowledge funding from the SERB-DST of the Government of India and from IIT Kanpur.

\section{Appendix: Additional data on devices with different interfaces}
Two additional multilayer MoS$_2$/STO FET devices were studied with different interfaces to verify the reproducibility of the counterclockwise hysteresis and the blocking transition of STO surface dipoles in such devices.
\begin{figure}
\centering
\includegraphics[width=3.4in]{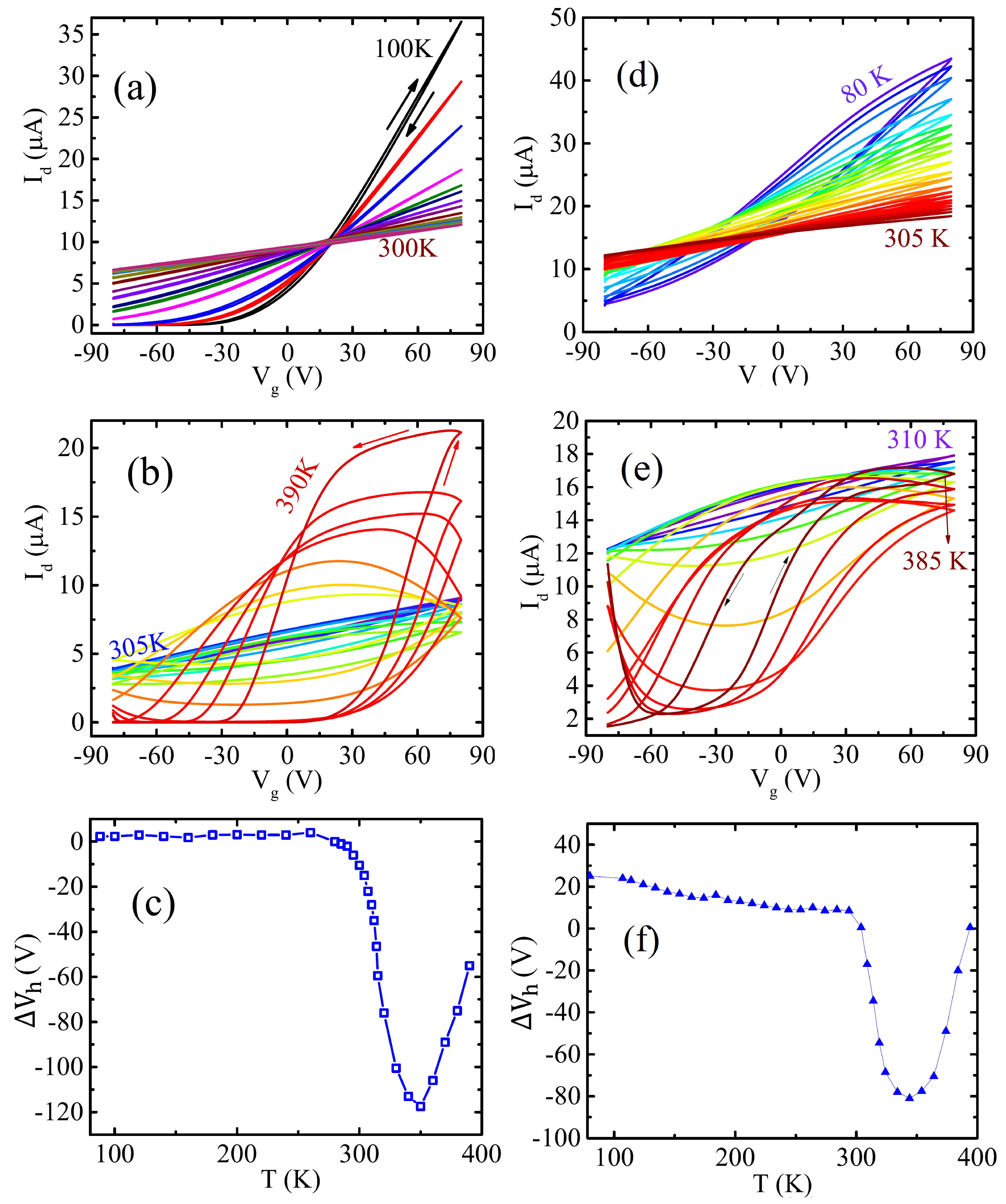}
\caption{(a) $I_{\rm d}-V_{\rm g}$ curves for different temperatures between 100 K and 300 K and (b)between 305 K and 390 K at $V_{\rm d}$ = 1V. (c) $\Delta V$ at different temperatures extracted from  the curves of (a) and (b). (a) $I_{\rm d}-V_{\rm g}$ curves for different temperatures between 80 K and 305 K and (b)between 310 K and 385 K at $V_{\rm d}$ = 1V. (c) $\Delta V$ at different temperatures extracted from  the curves of (a) and (b).}
\label{fig:sup}
\end{figure}

Figures \ref{fig:sup}(a,b) show the $I_{\rm d}-V_{\rm g}$ curves of an MoS$_2$/STO FET without HMDS passivation on the STO surface. This was measured at $V_{\rm d}=1$ V over a temperature range from 80 K to 390 K, using the same $V_{\rm g}$ sweep range and rate as described in the main paper. Figs. \ref{fig:sup}(d,e) show the same for an MoS$_2$ FET prepared on the rough surface of STO and with HMDS passivation and over a temperature range from 80 K to 395 K. Both these samples display a counterclockwise hysteresis at room temperature, which decreases sharply and vanishes upon cooling, as shown in Fig. \ref{fig:sup}(a) and (d). Upon heating, the hysteresis increases and peaks around 350 K before decreasing again, see Figs. \ref{fig:sup}(c) and (f). This latter sample indicates that surface roughness did not alter the polarization dynamics at the MoS$_2$-STO interface. This result contrasts with previous work on PZT-gated MoS$_2$ FETs \cite{counterclockwise1}, where surface roughness changed the hysteresis direction from anticlockwise to clockwise due to the interface states induced by defects on the rough surface. The below room temperature behavior of the three studied devices has differences as elaborated earlier but the blocking transition and the clockwise hysteresis above room temperature in all three studied devices are nearly the same.


\section{References}
\begin{thebibliography}:
\bibitem{TMDs}Z. Zhou, and Y. K. Yap, ``Two-dimensional electronics and optoelectronics: present and future Electronics", Electronics {\bf 6}, 53 (2017).
\bibitem{direct gap}K. F. Mak, C. Lee, J. Hone, J. Shan, and T. F. Heinz, ``Atomically thin MoS$_2$: a new direct-gap semiconductor", Phys. Rev. Lett. {\bf 105}, 136805 (2010).
\bibitem{direct gap1} B. Radisavljevic, A. Radenovic, J. Brivio, V. Giacometti, and A. Kis, ``Single-Layer MoS$_2$ Transistors", Nat. Nanotechnol. {\bf 6}, 147 (2011).
\bibitem{logic}L. M. Martinez, N. J. Pinto, C. H. Naylor, and A. T. Johnson, ``MoS$_2$ based dual input logic AND gate", AIP Adv. {\bf 6}, 125041 (2016).
\bibitem{logic1}S. Wachter, D. K. Polyushkin, O. Bethge, and T. Mueller, ``A microprocessor based on a two-dimensional semiconductor", Nat. Commun. {\bf 8}, 14948 (2017).
\bibitem{high frequency}D. Krasnozhon, D. Lembke, C. Nyffeler, Y. Leblebici, and A. Kis, ``MoS$_2$ Transistors Operating at Gigahertz Frequencies", Nano Lett. {\bf 14(10)}, 5905 (2014).
\bibitem{photodetectors}O. L. Sanchez, D. Lembke, M. Kayci, A. Radenovic, and A. Kis, ``Ultrasensitive photodetectors based on monolayer MoS$_2$", Nat. Nanotechnol. {\bf 8}, 497 (2013).
\bibitem{light}O. L. Sanchez, E. A. Llado, V. Koman, A. F. Morral, A. Radenovic, and A. Kis, ``Light generation and harvesting in a van der Waals heterostructure", ACS Nano {\bf 8}, 3042 (2014).
\bibitem{photo1}M. L. Tsai, S. H. Su, J. K. Chang, D. S. Tsai, C. H. Chen, C. I. Wu, L. J. Li, L. J.  Chen, and J. H. He, ``Mono-layer MoS$_2$ heterojunction solar cells", ACS Nano {\bf 8(8)}, 8317 (2014).
\bibitem{MoS2 2012}D. J. Late, B. Liu, H. S. S. R. Matte, V. P. Dravid, and C. N. R. Rao, ``Hysteresis in Single-Layer MoS$_2$ Field Effect Transistors", ACS Nano {\bf 6(6)}, 5635 (2012).
\bibitem{MoS2 20161}Y. Park, H. W. Baac, J. Heo, and G. Yoo, ``Thermally activated trap charges responsible for hysteresis in multilayer MoS$ _{2} $ field-effect transistors", Appl. Phys. Lett. {\bf 108}, 083102 (2016).
\bibitem{MoS2 20162} Y. Y. Illarionov, G. Rzepa, M. Waltl, T. Knobloch, A. Grill, M. M. Furchi, T. Mueller, and T. Grasser, ``The role of charge trapping in MoS$_2$/SiO$_2$ and MoS$_2$/hBN field-effect transistors", 2D Mater. {\bf 3}, 035004 (2016).
\bibitem{Mos2 2023}S. P. Jana, S. Gupta, and A. K. Gupta, ``Blocking transition of interface traps in MoS$_2$/SiO$_2$ field-effect transistors", Phys. Rev. B {\bf 108}, 195411 (2023).
\bibitem{Ta2O5} N. Mohta, R. K. Mech, S. Sanjay, R. Muralidharan, and D. N. Nath, ``Artificial Synapse Based on Back-Gated MoS$_2$ Field-Effect Transistor with High-k Ta$_2$O$_5$ Dielectrics", Phys. Status Solidi A {\bf 217}, 2000254 (2020).
\bibitem{counterclockwise}T. Li, L. Gao, H. Xie, L. Ye, W. Yang, Q. Liu, and K. Li, ``Intrinsic counterclockwise hysteresis in Mn-doped Pb(Zr,Ti)O$_3$ gated MoS$_2$ field effect transistors", Mater. Res. Express {\bf 5(6)}, 066308 (2018).
\bibitem{counterclockwise1} Z. Lu, C. Serrao, A. I. Khan, L. You, J. C. Wong, Y. Ye, H. Zhu, X. Zhang, and S. Salahuddin, ``Nonvolatile MoS$_2$ field effect transistors directly gated by single crystalline epitaxial ferroelectric", Appl. Phys. Lett. {\bf 111(2)}, 022902 (2017).
\bibitem{both}  K. L. Ganapathi, M. Rath, and M. S. R. Rao, ``Polarization induced switching in PZT back gated multilayer MoS$_2$ FETs for low power non-volatile memory", Semicond. Sci. Technol. {\bf 34}, 055016 (2019).
\bibitem{clockwise} A. Lipatov, P. Sharma, A. Gruverman, and A. Sinitskii, ``Optoelectrical Molybdenum Disulfide (MoS$_2$) Ferroelectric Memories", ACS Nano {\bf 9}, 8089 (2015).
\bibitem{clockwise1} X.-W. Zhang, D. Xie, J.-L. Xu, Y.-L. Sun, X. Li, C. Zhang, and R.-X. Dai, ``MoS$_2$ field-effect transistors with lead zirconate-titanate ferroelectric gating", IEEE Electron Device Lett. {\bf 36}, 784 (2015).
\bibitem{clockwise2} C. Zhou, X. Wang, S. Raju, Z. Lin, D. Villaroman, B. Huang, and Y. Chai, ``Low voltage and high ON/OFF ratio field-effect transistors based on CVD MoS$_2$ and ultra high-k gate dielectric PZT", Nanoscale {\bf 7}, 8695 (2015).
\bibitem{STO5} T. Sakudo, and H. Unoki, ``Dielectric Properties of SrTiO$_3$ at Low Temperatures", Phys. Rev. Lett. {\bf 26}, 851 (1971).
\bibitem{GSTO film} S. Saha, O. Kahya, M. Jaiswal, A. Srivastava, A. Annadi, J. Balakrishnan, A. Pachoud, C.-T. Toh, B.-H. Hong, J.-H. Ahn, T. Venkatesan, and B. Ozyilmaz, ``Unconventional Transport Through Graphene on SrTiO$_3$: A Plausible Effect of SrTiO$_3$ Phase-Transitions", Sci. Rep. {\bf 4}, 6173 (2015).
\bibitem{GSTO film1} K. T. Kang, H. Kang, J. Park, D. Suh, and W. S. Choi, ``Quantum Conductance Probing of Oxygen Vacancies in SrTiO$_3$ Epitaxial Thin Film Using Graphene", Adv. Mater. {\bf 29}, 1700071 (2017).
\bibitem{GSTO bulk} N. J. G. Couto, B. Sacepe, and A. F. Morpurgo, ``Transport Through Graphene on SrTiO$_3$", Phys. Rev. Lett. {\bf 107}, 1-5 (2011).
\bibitem{GSTO both} R. Sachs, Z. Lin, and J. Shi, ``Ferroelectric-Like SrTiO$_3$ Surface Dipoles Probed by Graphene", Sci. Rep. {\bf 4}, No. 3657 (2014).
\bibitem{GSTO bulk1} A. Sahoo, D. Nafday, T. Paul, R. Ruiter, A. Roy, M. Mostovoy, T. Banerjee, T. Saha-Dasgupta, and A. Ghosh, ``Out-of-Plane Interface Dipoles and Anti-Hysteresis in Graphene-Strontium Titanate Hybrid Transistor", npj 2D Mater. Appl. {\bf 2}, No. 9 (2018).
\bibitem{GSTO bulk4} S. Chen, X. Chen, E. A. Duijnstee, B. Sanyal, and T. Banerjee, ``Unveiling temperature-induced structural domains and movement of oxygen vacancies in SrTiO$_3$ with graphene", ACS Appl. Mater. Interfaces {\bf 12}, 52915-52921 (2020).
\bibitem{1973} A. D. Bruce, and R. A. Cowley, ``Lattice dynamics of strontium titanate: anharmonic interactions and structural phase transitions", J. Phys. C: Solid State Phys. {\bf 6}, 2422 (1973).
\bibitem{1976} R. Migoni, H. Bilz, and D. Bäuerle, ``Origin of Raman scattering and ferroelectricity in oxidic perovskites", Phys. Rev. Lett. {\bf 37}, 1155 (1976).
\bibitem{STO6} S. E. Rowley, L. J. Spalek, R. P. Smith, M. P. M. Dean, M. Itoh, J. F. Scott, G. G. Lonzarich, and S. S. Saxena, ``Ferroelectric quantum criticality", Nat. Phys. {\bf 10}, 367-372 (2014).
\bibitem{STO7} A. Müller, and H. Burkard, ``SrTiO$_3$: An intrinsic quantum paraelectric below 4 K", Phys. Rev. B {\bf 19}, 3593 (1979).
\bibitem{STO-SHG} B. Kang, ``Second-harmonic Generation of Treated-STO Surface", J. Kor. Cer. Soc., {\bf 49}, 142 (2012).
\bibitem{XYZ}A. C. Gomez, M. Buscema, R. Molenaar, V. Singh, L. Janssen, H. S. Van Der Zant, and G. A. Steele, ``Deterministic transfer of two-dimensional materials by all-dry viscoelastic stamping", 2D Mater. {\bf 1}, 011002 (2014).
\bibitem{HMDS} S. P. Jana, S. Shivangi, S. Gupta, and A. K. Gupta, ``Enhanced performance of MoS$_2$/SiO$_2$ field-effect transistors by hexamethyldisilazane (HMDS) encapsulation", Appl. Phys. Lett. {\bf 124}, 241101 (2024).
\bibitem{Raman}P. Chen, W. Xu, Y. Gao, J. H. Warner, and M. R. Castell, "Epitaxial Growth of Monolayer MoS\textsubscript{2} on SrTiO\textsubscript{3} Single Crystal Substrates for Applications in Nanoelectronics", ACS Appl. Nano Mater. {\bf 1}, 6976-6988 (2018).
\bibitem{anomalous lattice vibration}C. Lee, H. Yan, L. E. Brus, T. F. Heinz, J. Hone, and S. Ryu, ``Anomalous Lattice Vibrations of Single and Few-Layer MoS$ _{2} $", ACS Nano {\bf 4(5)}, 2695 (2010).
\bibitem{Petzelt} J. Petzelt, T. Ostapchuk, I. Gregora, I. Rychetský, S. Hoffmann-Eifert, A. V. Pronin, Y. Yuzyuk, B. P. Gorshunov, S. Kamba, V. Bovtun, and J. Pokorný, ``Dielectric, infrared, and Raman response of undoped SrTiO\textsubscript{3} ceramics: Evidence of polar grain boundaries", Phys. Rev. B {\bf 64}, 184111 (2001).
\bibitem{STO1} N. Bickel, G. Schmidt, K. Heinz, and K. Müller, ``Ferroelectric relaxation of the SrTiO\textsubscript{3} (100) surface", Phys. Rev. Lett. {\bf 62}, 2009 (1989).
\bibitem{STO2} J. Hanzig, M. Zschornak, F. Hanzig, E. Mehner, H. Stöcker, B. Abendroth, C. Röder, A. Talkenberger, G. Schreiber, D. Rafaja, and S. Gemming, ``igration-induced field-stabilized polar phase in strontium titanate single crystals at room temperature", Phys. Rev. B Condens. Matter Mater. Phys. {\bf 88}, 024104 (2013).
\bibitem{STO3}T. Leisegang, H. Stöcker, A. A. Levin, T. Weißbach, M. Zschornak, E. Gutmann, K. Rickers, S. Gemming, and D. C. Meyer, ``Switching Ti valence in SrTiO\textsubscript{3} by a dc electric field", Phys. Rev. Lett. {\bf 102}, 087601 (2009).
\bibitem{STO4} E. Heifets, E. A. Kotomin, and J. Maier, ``Semi-empirical simulations of surface relaxation for perovskite titanates", Surf. Sci. {\bf 462}, 19-35 (2000).
\bibitem{graphene-hyst}A. K. Singh, and A. K. Gupta, ``Reversible control of doping in graphene-on-SiO$_2$ by cooling under gate-voltage", J. Appl. Phys. {\bf 122}, 195305 (2017).

\end{thebibliography}
\end{document}